\documentclass{article}
\usepackage{spconf,amsmath,graphicx}
\usepackage{enumitem}
\usepackage{booktabs}
\usepackage[colorlinks, citecolor=blue]{hyperref}
\usepackage{bbding}
\usepackage{caption}
\hypersetup{hypertex=true,
            colorlinks=true,
            linkcolor=blue,
            anchorcolor=blue,
            citecolor=blue}
\setlist{nosep, leftmargin=14pt}
\usepackage{hyperref}
\usepackage{mwe} 

\title{COROLLA: An Efficient Multi-Modality Fusion Framework with Supervised Contrastive Learning for Glaucoma Grading}
\name{ Zhiyuan Cai$^{1\dagger}$, Li Lin$^{1,2\dagger}$, Huaqing He$^{1}$, Xiaoying Tang$^{1*}$
\thanks{$^\dagger$ These authors contributed equally.}
\thanks{
$^*$Corresponding author: Dr. Xiaoying Tang (\url{tangxy@sustech.edu.cn})
}}

\address{
Department of Electrical and Electronic Engineering,
Southern University of Science and Technology, \\ Shenzhen, China\\
Department of Electrical and Electronic Engineering,
The University of Hong Kong,\\Hong Kong SAR, China
}
\begin{document}
%
\maketitle
\begin{abstract}
Glaucoma is one of the ophthalmic diseases that may cause blindness, for which early detection and treatment are very important. Fundus images and optical coherence tomography (OCT) images are both widely-used modalities in diagnosing glaucoma. However, existing glaucoma grading approaches mainly utilize a single modality, ignoring the complementary information between fundus and OCT. In this paper, we propose an efficient multi-modality supervised contrastive learning framework, named COROLLA, for glaucoma grading. Through layer segmentation as well as thickness calculation and projection, retinal thickness maps are extracted from the original OCT volumes and used as a replacing modality, resulting in more efficient calculations with less memory usage. Given the high structure and distribution similarities across medical image samples, we employ supervised contrastive learning to increase our models' discriminative power with better convergence. Moreover, feature-level fusion of paired fundus image and thickness map is conducted for enhanced diagnosis accuracy. On the GAMMA dataset, our COROLLA framework achieves overwhelming glaucoma grading performance compared to state-of-the-art methods.
\end{abstract}
\begin{keywords}
Supervised contrastive learning, multi-modality fusion, OCT, retinal thickness map, glaucoma
\end{keywords}
\vspace{1pt}
\section{Introduction}
\label{sec:intro}
\vspace{1pt}
Glaucoma is one of the ophthalmic diseases that cause damages to the optic nerve, resulting in potential vision loss \cite{7566373}. Glaucoma is considered as the second leading cause of blindness globally. Retinal fundus image is an inexpensive and readily available modality for eye examination. Optical coherence tomography (OCT) is a new 3D imaging technique for human tissues, especially for eye. There have been researches employing either fundus images \cite{5627290,9410772} or OCT images \cite{WU2012815,GARCIA2021102132} for diagnosing glaucoma. However, there is no such research jointly making use of both the two modalities.

\begin{figure}[tb]
\centering
     \includegraphics[width=9cm]{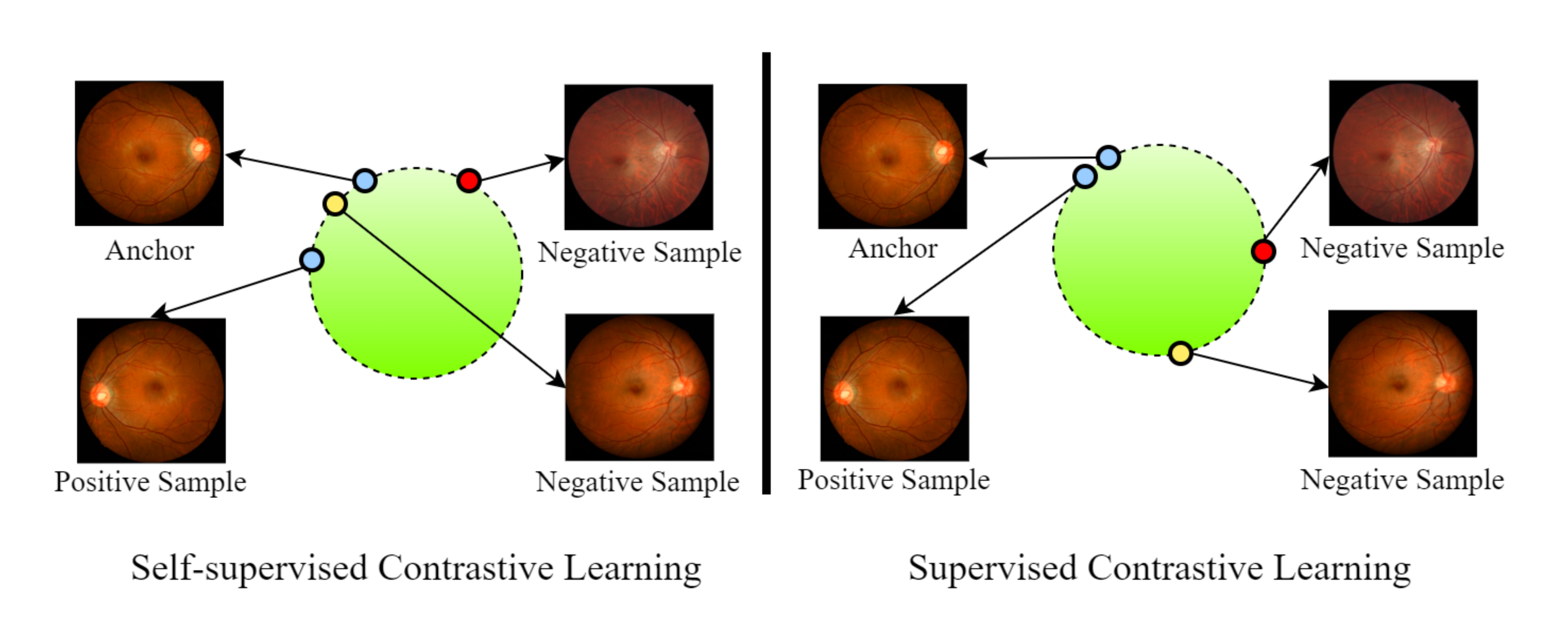}
     \caption{Self-supervised contrastive learning versus supervised contrastive learning in fundus images' embedding space.}
     \label{ssclandscl}
\end{figure}



Multi-modality is very common in medical imaging, under the assumption that they can provide complementary information for a specific task of interest \cite{ZHOU2019100004}. Multi-modality fusion is a popular research topic in medical imaging, including input-level fusion, feature-level fusion, and decision-level fusion \cite{111111}. Typically, a combination of multiple modalities provides more accurate results than each single modality \cite{ZHOU2019100004}. 



Self-supervised contrastive learning has flourished in recent years, leading to state-of-the-art (SOTA) performance in unsupervised training of deep image models. Self-supervised contrastive learning methods pull together positive samples and an anchor while push apart negative samples from the anchor in an embedding space. However, since there are high structure and distribution similarities across medical image samples (as shown in Fig. \ref{ssclandscl}), anchors from different classes are similar, which may make self-supervised contrastive learning fail. Later, \cite{khosla2021supervised} extends self-supervised batch contrastive to the fully-supervised setting, named as supervised contrastive learning. Supervised contrastive learning leverages label information, and thus can naturally address the aforementioned anchor similarity issue.

In such context, we propose an efficient multi-modality fusion framework with supervised contrastive learning for glaucoma grading, named as COROLLA. The contributions of our proposed COROLLA are three-fold: (1) Based on clinical prior knowledge, we project and compress OCT images into thickness maps which require less GPU-memory consumption and perform better in glaucoma grading. (2) COROLLA takes advantages of supervised contrastive learning and performs outstandingly in glaucoma grading. (3) Multi-modality feature-level fusion is employed in COROLLA to enhance the grading performance, and the feasibility and effectiveness
of COROLLA have been established through extensive experiments. The source codes are available at \url{https://github.com/Davidczy/SupCon_GAMMA}.
\section{Methods}
\vspace{2.0pt}
\label{sec:pagestyle}

\vspace{1pt}
\subsection{Retinal thickness map generation}
\vspace{1pt}

OCT can characterize tissue thickness and distance provided by the interface reflection of different tissues. However, in deep learning methods, utilizing raw OCT volumes would consume a lot of memory and time. Resizing can effectively reduce the consumption of computing resources, but resized images may lose a lot of critical image information. Therefore, a compressed representation with clinical prior is needed to more efficiently make use of OCT volumes.


\begin{figure}[htb]
\centering
     \includegraphics[width=8.5cm]{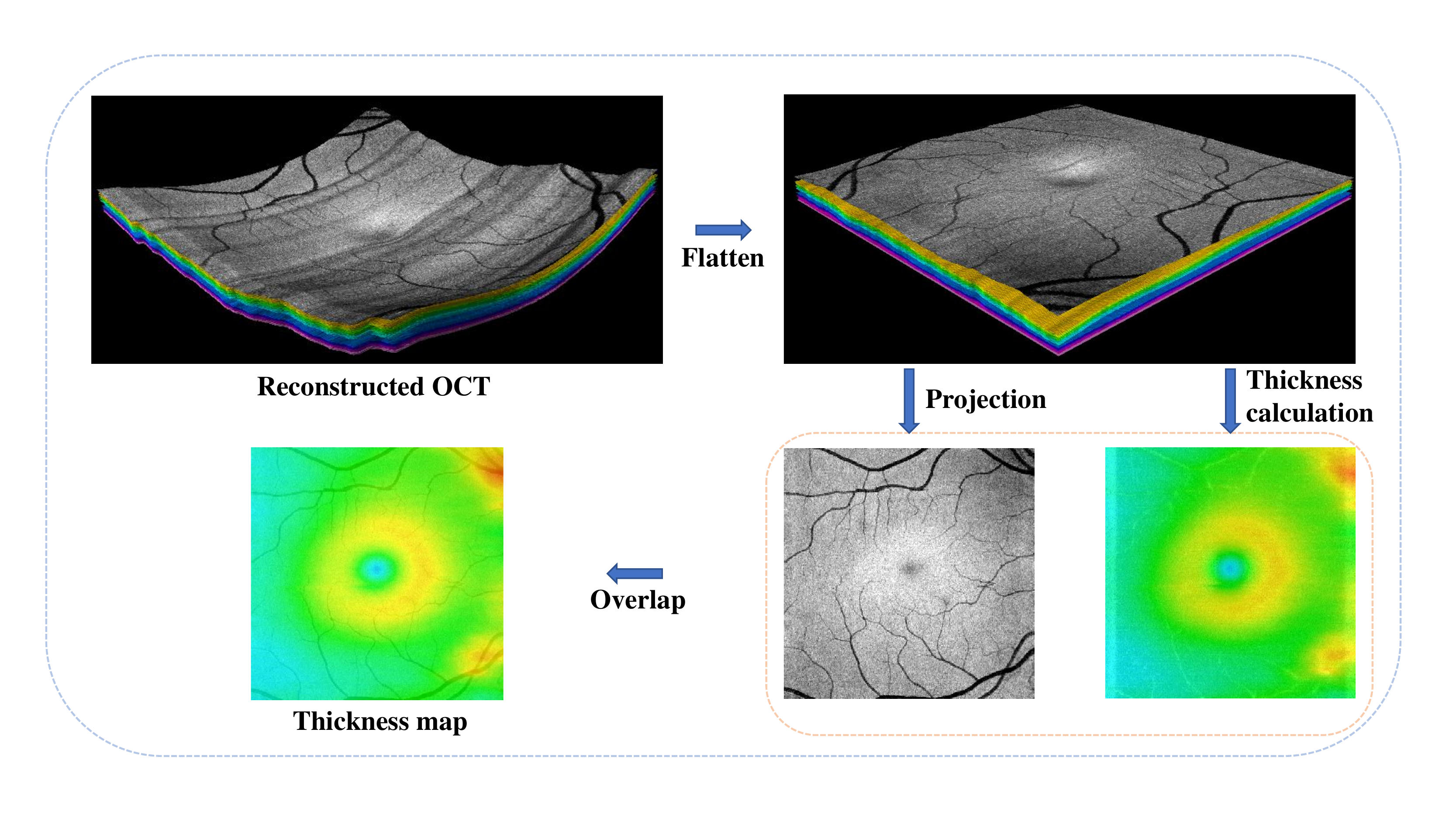}
     \caption{The thickness map generation procedure.}
     \label{generation}
\end{figure}
We use the software OCTExplorer \cite{4799172} to transform OCT volumes to thickness maps (Fig. \ref{generation}), which involves the following steps:
\begin{enumerate}[]
    \item Reconstruct the OCT volume with 256 slices.
    \item Identify non-intersecting surfaces between different layers. Voxels between adjacent surfaces belong to the same layer.
    \item Define a cost function related to different surfaces.
    \item Flatten the 3D OCT volume to provide a more consistent shape for a segmentation purpose.
    \item Minimize the cost function to identify feasible solutions of the flattened surfaces. 
    \item Calculate thickness between different layers. 
    \item Map thickness to RGB by \textit{turbo} \cite{turbo} colormap.
    \item Project the nerve fiber layer and fanglion cell layer to a common plane.
    \item Overlay the projected image onto the RGB thickness image to obtain the final thickness map.
\end{enumerate}

In terms of size, the resized thickness map is $512 \times 512$ while the raw OCT volume is $256 \times 512 \times 992$. It is obvious that such volume-to-thickness projection can significantly reduce the GPU-memory consumption. In later experiments, we will also show that the thickness map is more discriminative than the raw OCT volume for glaucoma grading.
\vspace{1pt}
\subsection{COROLLA: An efficient classification framework}
\vspace{1pt}
\begin{figure}[htb]
\centering
     \includegraphics[width=8.5cm]{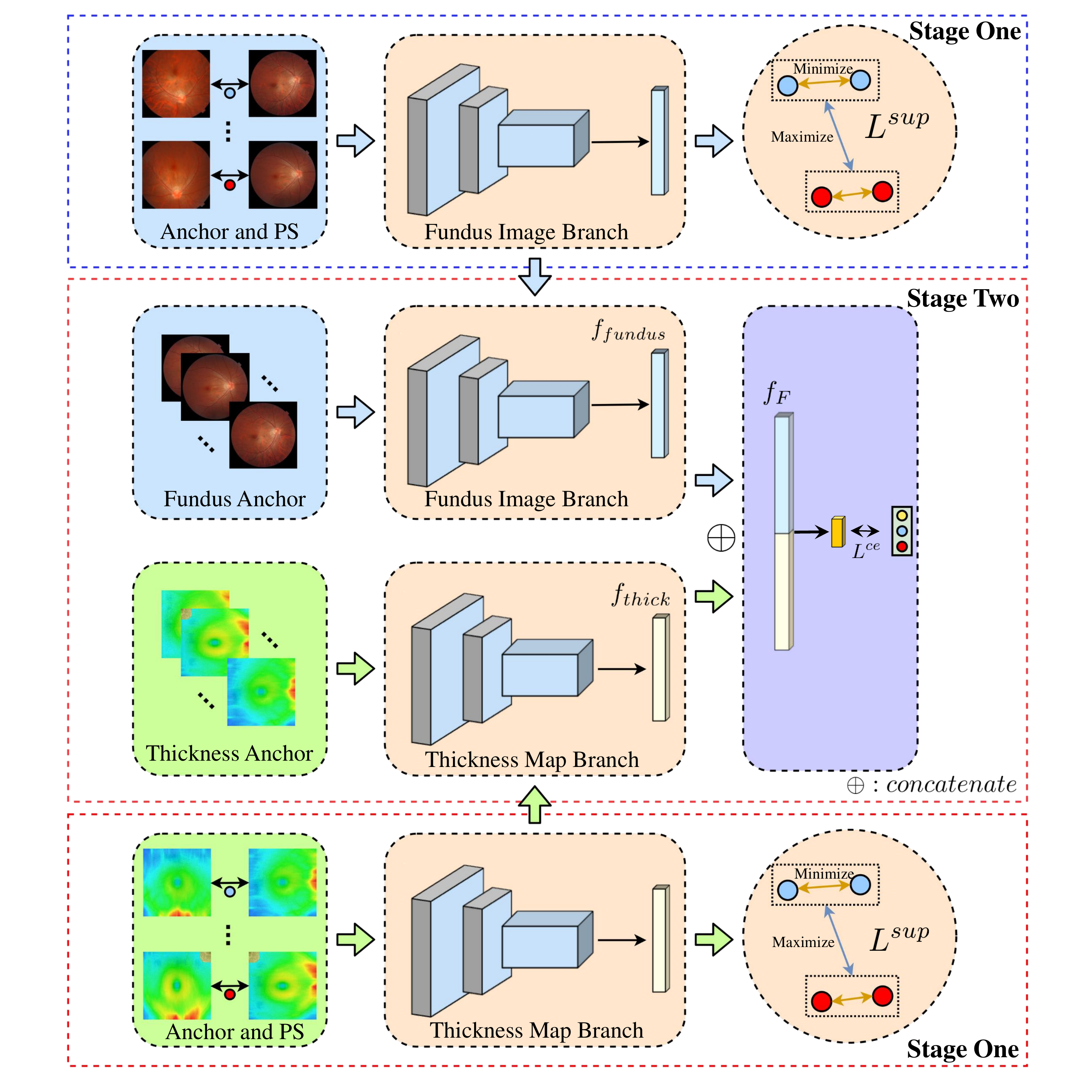}
     \caption{The overall pipeline of COROLLA. Keys: PS -- positive sample.}
     \label{pipeline}
\end{figure}

As shown in Fig. \ref{pipeline}, our COROLLA is a two-stage framework. In stage one, we separately extract features from the retinal fundus image branch and the thickness map branch with encoders of the same structure. Later, we train both encoders through supervised comparative learning \cite{khosla2021supervised} to make the features more discriminative, which is beneficial for the subsequent classification task.

In stage two, we again extract features from the aforementioned two branches pretrained in stage one. The outputs of the two branches are features of the fundus images and the OCT images, annotated as $f_{fundus}$ and $f_{thick}$. We concatenate the features together to be our finally-acquired feature set $f_F$. We then feed $f_F$ into the classification layer to predict the glaucoma grade.
\vspace{0.1pt}
\subsubsection{Supervised contrastive learning}
\vspace{1pt}
The supervised contrastive loss is developed from the self-supervised contrastive loss. Given a set of $n$ samples, denoted as $A$, anchor $x_i$ is the $i^{th}$ member in $A$. Anchor $x_i$'s positive sample, which is an augmentation from $x_i$, is denoted as $z_i$. The self-supervised contrastive learning loss is

\begin{equation} \label{selfconloss}
    L^{se} = \sum_{i \in I} L_i^{se} = -\sum_{i \in I}log \frac{exp(z_i\cdot z_{j(i)}/\tau)}{\sum_{a\in A(i)}exp(z_i \cdot z_a /\tau)},
\end{equation}
where $\cdot$ denotes the inner dot product, $\tau \in R^+$ is a scalar temperature parameter, $A(i) = A \backslash \{x_i\}$.

For supervised contrastive learning, the contrastive loss in Eq. \ref{selfconloss} needs to be modified, taking the following form \cite{khosla2021supervised}

\begin{equation}
    \begin{aligned}
        L^{sup} &= \sum_{i \in I} L_i^{sup} \\
        &= \sum_{i \in I}\frac{-1}{|P(k)|}\sum_{p \in P(k)}log \frac{exp(z_i\cdot z_{j(i)}/\tau)}{\sum_{a\in A(i)}exp(z_i \cdot z_a /\tau)},
    \end{aligned}
\end{equation}
where $P(k)$ is the set of indices in class $k$, and $|P(k)|$ is its cardinality. We use $L^{sup}$ in stage one to leverage label information, attending to achieve better convergence.
\vspace{1pt}
\subsubsection{Multi-modality fusion}
\vspace{1pt}
After supervised contrastive learning, the fundus image branch and the thickness map branch will generate two sets of discriminative features. It is worth pointing out that these feature representations contain complementary information as well as label information. Hence, under such circumstances, feature fusion can maximize the advantages of the two sets of features. So we concatenate the features together and feed them into a subsequent fully-connected layer. 

\vspace{3.0pt}
\section{Experiments and results}
\vspace{2.0pt}
\label{sec:typestyle}
\vspace{1pt}
\subsection{Dataset}
\vspace{1pt}
The GAMMA dataset consists of 100 paired fundus images and OCT images with three-level glaucoma grading (50 “None”, 26 “Early”, and 24 “Mid-Advanced”) labeled by ophthalmologists. Each OCT volume contains 256 2D slices from the B channel. Five-fold cross-validation experiments are conducted.
\vspace{1pt}
\subsection{Implementation}
\vspace{1pt}
The entire pipeline is implemented by PyTorch on a workstation equipped with NVIDIA RTX TITAN GPUs. 

In the training phase of stage one, we use OCTExplorer to transform OCT images into thickness maps. Afterwards, we resize each fundus image and each thickness map to $1024 \times 1024$ and $384 \times 384$. The feature extractor in each branch is ResNet50 pretrained on ImageNet. We use two fully-connected layers with ReLU to generate a 128-D tensor used for the loss $L^{sup}$. We set the batch size to be 8 and use Adam as our optimizer with an initial learning rate of 1e-3 and cosine decay. The network is trained for 10 more epochs after convergence. We fix the temperature $\tau$ in $L^{sup}$ as 0.05. For the augmentation method used to generate positive samples, we use a combination of random color jitter, random grayscaling, random center cropping and random horizontal flipping in the fundus image branch as well as a combination of random center cropping and random horizontal flipping in the thickness map branch.

In the training phase of stage two, we use the model pretrained in stage one. We remove the final fully-connected layers in both branches. Then, we perform feature-level fusion by concatenating the two set of features from the two branches. Afterwards, we feed the concatenated features to a fully-connected layer, with an input size of 4096 and an output size of 3. In this process, we again resize each fundus image and each thickness map to $1024 \times 1024$ and $384 \times 384$. We set the batch size to be 8 and use Adam as our optimizer with an initial learning rate of 2e-3. We adopt 1000 more epochs after convergence. We use cross entropy to be our loss $L^{ce}$.

\vspace{1pt}
\subsection{Evaluation}
\vspace{1pt}
We employ two quantitive evaluation metrics, namely accuracy (Acc) and cohen's kappa coefficient (Kappa) \cite{kappa}. Given the distribution of our samples in each category is uneven, we use Kappa since it accommodates well evaluations with unbalanced datasets.
\vspace{1pt}
\subsection{Ablation Experiments and Results}
\vspace{1pt}
\begin{table}[]
\centering
\caption{Ablation analysis results of our proposed COROLLA on glaucoma grading, in terms of Acc and Kappa. SCL stands for supervised contrastive learning.}
\begin{tabular}{@{}ccccc@{}}
\toprule

Fundus &Thickness    & SCL& Acc & Kappa \\ \midrule
\Checkmark& &               & 0.830           & 0.728        \\
  &\Checkmark&            & 0.690           & 0.634        \\
  \Checkmark  &  \Checkmark&            & 0.840           & 0.741       \\
\Checkmark& &\Checkmark               & 0.880           & 0.806       \\

  &\Checkmark&\Checkmark               & 0.770           & 0.635      \\

 \Checkmark&\Checkmark&\Checkmark             & \textbf{0.900}           & \textbf{0.855}        \\ \bottomrule
\end{tabular}
\label{SCL}
\end{table}

\begin{table}[] 
\centering
\caption{Quantitative comparisons of using the original OCT image and the thickness map for glaucoma grading, in terms of Kappa, parameters and training time.}
\begin{tabular}{@{}cccc@{}}
\toprule
Modality + Model & Kappa &  Param &Time (s) \\ \midrule
OCT + 2D-ResNet & 0.606 & 23.5M&20,280 \\
OCT + 3D-ResNet & 0.652 & 46.9M & 60,685\\
Thickness + 2D-ResNet & \textbf{0.743}  & \textbf{23.5M}& \textbf{6,935}\\ 
\bottomrule
\end{tabular}
\label{thickoct}
\end{table}

In Table \ref{SCL}, we conduct two sets of experiments under two settings. Specifically, in the first setting, we separately feed fundus images, thickness maps, and both of them into COROLLA without supervised contrastive learning. In the second setting, we use supervised contrastive learning to train each branch first, and then repeat the same procedure as in the first setting. The quantitative results demonstrate that both feature fusion and supervised contrastive learning can improve the glaucoma grading performance. 


In Table \ref{thickoct}, we compare the performance and efficiency between OCT volume and thickness map. To be specific, we compare three different combinations: thickness map with 2D-ResNet, OCT image with 2D-ResNet and OCT image with 3D-ResNet. In OCT image with 2D-ResNet, we feed an OCT volume into ResNet as a 256-channel 2D image. Compared with the original OCT image, we find that the thickness map not only occupies less time and GPU-memory consumption, but also yields a better classification result. 


\begin{table}[] 
\centering
\caption{Comparisons between COROLLA and SOTA methods on glaucoma grading.}
\begin{tabular}{@{}cccc@{}}
\toprule
Methods    &Dataset    &Acc& Kappa \\ \midrule
Gabriel et al. \cite{GARCIA2021102132}&private&0.818&0.719\\
Parashar et al. \cite{9296796}&HRF \cite{HRF}&0.863&-\\
Cheng et al. \cite{6091111}&private \cite{5627290} &0.798 & 0.641\\
COROLLA & GAMMA&\textbf{0.900} &\textbf{0.855}    \\ \bottomrule
\end{tabular}
\label{SOTA}
\end{table}

In Table \ref{SOTA}, we compare our COROLLA with SOTA glaucoma grading methods. Gabriel et al. \cite{GARCIA2021102132} and Cheng et al. \cite{6091111} evaluate their proposed models on other glaucoma datasets with more than 1000 samples. Parashar et al. \cite{9296796} employs a two-stage model to grade glaucoma on the HRF \cite{HRF} dataset. Our proposed COROLLA is much better than all three compared methods, in terms of both accuracy and Kappa. However, please note the comparisons may not be fair enough since these four sets of results are reported based on different datasets. The GAMMA dataset is recently released, and thus there are no reported results on GAMMA that we can compare with.

\section{Conclusion}
\vspace{4.0pt}
In this work, we proposed and validated an efficient multi-modality fusion framework with supervised contrastive learning. In this framework, we successfully identified the effectiveness of supervised contrastive learning and feature-level multi-modality fusion. In addition, we observed that a thickness map worked better than its original OCT volume, in terms of both classification accuracy and computational efficiency, on the glaucoma grading task. A potential limitation of this work is that the size of our GAMMA dataset is relatively limited. Techniques in the realm of few-shot learning may better accommodate the GAMMA dataset, which is a research topic that we aim to explore in our future research work.

\label{sec:print}

\section{acknowledgements}
This study was supported by the Shenzhen Basic Research Program (JCYJ20190809120205578); the National Natural Science Foundation of China (62071210); the Shenzhen Basic Research Program (JCYJ20200925153847004); the High-level University Fund (G02236002).
\label{sec:page}


\bibliographystyle{IEEEbib}
\bibliography{strings,refs,ref}

\begin{thebibliography}{10}
\scriptsize

\bibitem{7566373}
Sharanagouda Nawaldgi,
\newblock ``Review of automated glaucoma detection techniques,''
\newblock in {\em 2016 International Conference on Wireless Communications,
  Signal Processing and Networking (WiSPNET)}, 2016, pp. 1435--1438.

\bibitem{5627290}
Cheng et~al.,
\newblock ``Closed angle glaucoma detection in retcam images,''
\newblock in {\em 2010 Annual International Conference of the IEEE Engineering
  in Medicine and Biology}, 2010, pp. 4096--4099.

\bibitem{9410772}
Siva Raja~P. M et~al.,
\newblock ``Automatic glaucoma diagnosis based on photo segmentation with
  fundus images,''
\newblock in {\em 2021 International Conference on Computational Intelligence
  and Knowledge Economy (ICCIKE)}, 2021, pp. 102--105.

\bibitem{WU2012815}
Huijuan Wu et~al.,
\newblock ``Diagnostic capability of spectral-domain optical coherence
  tomography for glaucoma,''
\newblock {\em American Journal of Ophthalmology}, vol. 153, no. 5, pp.
  815--826.e2, 2012.

\bibitem{GARCIA2021102132}
Gabriel García et~al.,
\newblock ``Circumpapillary oct-focused hybrid learning for glaucoma grading
  using tailored prototypical neural networks,''
\newblock {\em Artificial Intelligence in Medicine}, vol. 118, pp. 102132,
  2021.

\bibitem{ZHOU2019100004}
Tongxue Zhou et~al.,
\newblock ``A review: Deep learning for medical image segmentation using
  multi-modality fusion,''
\newblock {\em Array}, vol. 3-4, pp. 100004, 2019.

\bibitem{111111}
Lior Rokach,
\newblock ``Ensemble-based classifiers,''
\newblock {\em Artificial Intelligence Review}, vol. 33, no. 39, 2010.

\bibitem{khosla2021supervised}
Prannay Khosla et~al.,
\newblock ``Supervised contrastive learning,'' 2021.

\bibitem{4799172}
Garvin et~al.,
\newblock ``Automated 3-d intraretinal layer segmentation of macular
  spectral-domain optical coherence tomography images,''
\newblock {\em IEEE Transactions on Medical Imaging}, vol. 28, no. 9, pp.
  1436--1447, 2009.

\bibitem{turbo}
Anton Mikhailov,
\newblock ``Turbo, an improved rainbow colormap for visualization,''
\newblock 2019.

\bibitem{kappa}
Marry~L. McHugh,
\newblock ``Interrater reliability: the kappa statistic,''
\newblock {\em Biochemia Medica}, vol. 22, no. 3, pp. 276--282, 2012.

\bibitem{9296796}
Parashar et~al.,
\newblock ``Automatic classification of glaucoma stages using two-dimensional
  tensor empirical wavelet transform,''
\newblock {\em IEEE Signal Processing Letters}, vol. 28, pp. 66--70, 2021.

\bibitem{HRF}
Odstrcilik et~al.,
\newblock ``Retinal vessel segmentation by improved matched filtering:
  Evaluation on a new high-resolution fundus image database,''
\newblock {\em Image Processing, IET}, vol. 7, pp. 373--383, 06 2013.

\bibitem{6091111}
Cheng et~al.,
\newblock ``Focal edge association to glaucoma diagnosis,''
\newblock in {\em 2011 Annual International Conference of the IEEE Engineering
  in Medicine and Biology Society}, 2011, pp. 4481--4484.

\end{thebibliography}

\end{document}